\documentclass[aps, prd, twocolumn, superscriptaddress, floatfix,
               nofootinbib, longbibliography]{revtex4-1}

\usepackage{amsmath}
\usepackage{amssymb}

\usepackage[]{graphicx}
\usepackage{subfig}

\usepackage{color}

\usepackage{natbib}

\usepackage{placeins}

\usepackage{multirow}

\usepackage{hhline}

\usepackage{hyperref}
\hypersetup{colorlinks, allcolors=blue!60!black}

\usepackage{siunitx}

\newcommand{\fdot}{\dot{f}}
\newcommand{\F}{{\mathcal{F}}}
\newcommand{\twoFhat}{\widehat{2\F}}
\newcommand{\twoFtilde}{\widetilde{2\F}}
\newcommand{\A}{\boldsymbol{\mathcal{A}}}
\newcommand{\blambda}{\boldsymbol{\mathbf{\lambda}}}

\newcommand{\tglitch}{t^{\rm g}}
\newcommand{\tstart}{t^{s}}
\newcommand{\tend}{t^{e}}
\newcommand{\Ng}{N_{\rm g}}

\newcommand{\tref}{t_{\rm ref}}
\newcommand{\Ngrid}{M}
\newcommand{\Nseg}{N}

\renewcommand{\H}{\mathcal{H}}
\newcommand{\Hs}{\H_{\rm S}}
\newcommand{\Hn}{\H_{\rm N}}
\newcommand{\Hg}{\H_{\rm gS}}

\newcommand{\Bsn}{B_{\rm S/N}}
\newcommand{\Bgn}{B_{\rm gS/N}}
\newcommand{\Bgs}{B_{\rm gS/S}}

\newcommand{\smax}{{s_{\textrm{max}}}}

\newcommand{\rhohatmax}{\hat{\rho}_{\mathrm{max}}}

\newcommand{\N}{\mathcal{N}}
\newcommand{\fk}{f^{(k)}}
\newcommand{\data}{\boldsymbol{x}}
\newcommand{\sky}{\boldsymbol{\Omega}}

\newcommand{\ddl}{\{\!\!\{}
\newcommand{\ddr}{\}\!\!\}}
\newcommand{\inject}{{\mathrm{inject}}}

\usepackage[normalem]{ulem}	
\usepackage[usenames,dvipsnames]{xcolor}

\begin{document}

\title{A semicoherent glitch-robust continuous gravitational wave search}

    \author{G. Ashton}
    \email[E-mail: ]{gregory.ashton@ligo.org}
    \affiliation{Max Planck Institut f{\"u}r Gravitationsphysik
                 (Albert Einstein Institut) and Leibniz Universit\"at Hannover,
                 30161 Hannover, Germany}
    \affiliation{Monash Centre for Astrophysics,
                 School of Physics and Astronomy,
                 Monash University, VIC 3800, Australia}
    \author{R. Prix}
    \affiliation{Max Planck Institut f{\"u}r Gravitationsphysik
                 (Albert Einstein Institut) and Leibniz Universit\"at Hannover,
                 30161 Hannover, Germany}
    \author{D.I. Jones}
    \affiliation{Mathematical Sciences, University of Southampton,
                 Southampton SO17 1BJ, UK}

\date{\today}

\begin{abstract}

Continuous gravitational-wave signals from isolated non-axisymmetric rotating
neutron stars may undergo episodic spin-up events known as glitches.
 If unmodelled by a search, these can result in missed or
misidentified detections. We outline a semicoherent glitch-robust search
method that allows identification of glitching signal candidates and inference
about the model parameters.

\end{abstract}

\pacs{04.80.Nn, 97.60.Jd, 04.30.Db}

\maketitle

\section{Introduction}

Continuous gravitational wave (CW) searches for rotating neutron stars
typically assume an underlying signal model (a \emph{template}) for the signal
observed in the detector and then perform a matched-filter analysis (see, e.g., \citet{abbott2017allsky, abbott2017allskyeinstein}).  These
templates assume the phase evolution of the source is well modelled by a spin
frequency, and several frequency derivatives. On the contrary, observations of
pulsars demonstrate that neutron stars are subject to low frequency
\emph{timing noise} \citep{hobbs2010} and can also undergo sudden spontaneous
increases in their rotation frequency and frequency derivatives known as
\emph{glitches} \citep{espinoza2011,
fuentes2017}. While the former effect is unlikely to have a substantial
negative impact for searches of data lasting less than a year \citep{jones2004,
ashton2015}, typical glitches seen in the pulsar population
may adversely affect current and
ongoing CW searches. In \citet{ashton2017}, we provide a statistical analysis
of pulsar glitches and demonstrated that for a fully
coherent matched-filter analysis, a glitch can cause a substantial relative
loss of signal-to-noise ratio (SNR); semicoherent searches (in which the data
is segmented, searched coherently and then recombined; for a review see
\citet{prix2009gravitational}) will suffer smaller
relative losses of SNR by comparison, but, during the follow-up
process\footnote{A follow-up refers to the process whereby a candidate from a
semicoherent search is subjected to a series of searches, each of which
increases the coherence time until the candidate is detected with a
fully coherent search}, a glitching candidate's SNR will not increase as expected
potentially resulting in dismissal of the candidate.

In this work, we introduce a \emph{glitch-robust} detection statistic in which
the template also models the size and epoch of one or more glitches.
Standard CW searches (by which we mean those using a non-glitch-robust detection statistic)
are already computationally constrained, so adding additional parameters
will increase the computational load and also reduce the significance of
results due to the increased number of trials. Moreover, wide-parameter space
searches typically begin with a semicoherent stage, which, as previously
mentioned, is more robust to glitches. Therefore, we do not advise modifying
the initial semicoherent blind search strategies, but that during the
follow-up and vetting of candidates, a glitch-robust statistic be used to guard
against dismissal of glitching signals.

We begin in Sec.~\ref{sec_glitch_robust} by defining a glitch-robust
detection statistic. Then in Sec.~\ref{sec_pe} we give a discussion and
comparison of how the statistic could be applied in a grid- or Markov chain Monte Carlo (MCMC)-based
search for CW candidates. In Sec.~\ref{sec_bayes_factor} we discuss how to
perform a model selection between glitching and non-glitching signals. Since
most glitching candidates will initially be identified by a standard-CW search, in
Sec.~\ref{sec_standard} we discuss how glitching signals might manifest
in such searches and what simple steps can be taken to identify them.
We conclude with overall discussion in Sec.~\ref{sec_discussion}.

\section{Semi-coherent glitch-robust detection}
\label{sec_glitch_robust}

In this section, we introduce the glitch-robust detection statistic, an
adaptation of the semicoherent $\F$-statistic for glitching signals. We
begin by defining the standard-CW $\F$-statistic and then describe the
glitch-robust modification.

For an isolated CW signal, the gravitational wave signal template,
$h(t)$, has two sets of parameters: the \emph{amplitude parameters} $\A=\{h_0,
\cos\iota, \psi, \phi_0\}$, consisting of the CW amplitude $h_0$,
inclination angle $\iota$, polarization angle $\psi$ and initial phase
$\phi_0$, and the \emph{phase-evolution} parameters
$\blambda=\{\sky, f, \dot{f}, \ldots\}$, consisting of the sky location
$\sky$, frequency $f$ and higher-order frequency derivatives $\fk$
(cf.~\citet{prix2009gravitational} for a general review).
One key component of defining $h(t)$ is the \emph{source-frame phase evolution},
which for a standard-CW signal can be written as (e.g., see Eq.~(18) of \citet{jks1998})
\begin{align}
\varphi(t) = 2\pi\sum_{k=0}^\smax \fk \frac{(t-\tref)^{k+1}}{(k+1)!}\,,
\end{align}
where $\tref$ is a reference time, $\fk$ is the $k$th frequency
derivative and $\smax$ is the number of spin-downs included in the template.

In this work, we model the $\ell$th glitch by $\delta\fk_\ell$, the
permanent increment in the
$k$th frequency derivative at an epoch $\tglitch_\ell$. For a source with
$\Ng$ glitches, the glitching source phase evolution is then
\begin{align}
\varphi'(t) = \varphi(t)
+ 2\pi
\sum_{\ell=0}^{\Ng}
H(t-\tglitch_\ell)
\sum_{k=0}^{\smax}
\delta \fk_\ell \frac{(t-\tglitch_\ell)^{k+1}}{(k+1)!}\,,
\label{eqn_glitching_spe}
\end{align}
where
$H(t)$ is the unit
step function. This is analogous to the method used in pulsar timing
\citep{edwards2006}, except that we do not model any exponentially
decaying components.

We refer to $\{\fk\}$ as the set of the frequency
and its derivatives up to $\smax$, $\ddl \delta \fk_{\ell}\ddr$ as the set of
all glitch magnitudes for all glitches, and $\{\tglitch_\ell\}$ as the set of all glitch epochs.

The fully coherent $\F$-statistic,
used by many wide-parameter space searches as a ranking statistic, is
the log-likelihood ratio for signal vs.\ Gaussian noise, marginalized over the amplitude
parameters \citep{jks1998, prix2009, prix2011}. Using only data spanning
times $[\tstart, \tend]$ from the full set of data $\data$, we write the
fully coherent statistic as $\twoFtilde(\data; \blambda, \tstart, \tend)$.
Often, wide-parameter space searches use a semicoherent approach in which
the total data span $T$ of $\data$ is divided into $\Nseg$ contiguous
\emph{segments}. Defining $\{t_\ell\}$ as the set of start times for
each segment, the semicoherent $\F$-statistic is
\begin{align}
\begin{split}
\twoFhat\left(\data; \blambda, \Nseg\right)
\equiv \sum_{\ell=1}^{\Nseg} \twoFtilde\left(
\data; \blambda, t_\ell, t_\ell{+}\frac{T}{\Nseg}\right)\,.
\end{split}
\label{eqn_semi}
\end{align}

Ideally, a glitch-robust statistic would modify the standard-CW
fully coherent $\F$-statistic with the glitching source phase evolution,
Eq.~\eqref{eqn_glitching_spe}, resulting in a fully coherent glitch-robust
detection statistic.

However, we propose instead the following pragmatic approach: let us use a
semicoherent detection statistic with the glitch-epochs $\tglitch_\ell$
partitioning \emph{segments}. Then defining $\twoFtilde\left(\data; \blambda,
\{\delta \fk_\ell\}, \tglitch_\ell, \tglitch_{\ell+1}\right)$ as the
fully coherent detection statistic calculated between glitches and assuming the
source phase model of Eq.~\eqref{eqn_glitching_spe}, we can define a
glitch-robust semicoherent $\mathcal{F}$-statistic:
\begin{align}
\twoFhat\!\left(\!\data; \blambda, \ddl \delta \fk_{\ell}\ddr,
 \{\tglitch_\ell\}\!\right) \!
\equiv \!
\sum_{\ell=0}^{\Ng}
\twoFtilde\left(\data; \blambda, \{\delta \fk_\ell\},
\tglitch_\ell, \tglitch_{\ell+1}\right)\,.
\label{eqn_glitch_likelihood}
\end{align}
For convenience, we also define $\tglitch_0$ and $\tglitch_{\Ng+1}$ to
coincide with the start and end time of the data used.

In this semicoherent detection statistic there are $\Ng+1$ contiguous segments
which is implied by the size of $\{\tglitch_\ell\}$, with the first glitch occurring
at $\tglitch_{\ell=1}$.

This simplistic method leverages readily available and tested code. However,
this approach is potentially sub-optimal compared to a fully coherent glitch-robust
detection statistic. By using the semicoherent statistic
over glitches, we allow for independent amplitude parameters $\A$ in
each inter-glitch segment. This allows not only for a jump in phase,
which would be physically plausible for a glitch, but also in
amplitude $h_0$ and polarization angles $\cos\iota,\psi$, which is more
likely to be unphysical.
In principle one could build a coherent glitch-robust statistic by
allowing only a phase jump in each glitch as an additional search
parameter to the $\fk$-jumps, but it is unclear if this would gain
much extra sensitivity, and we postpone this to a future study.

\section{Glitch-robust searches and parameter estimation}
\label{sec_pe}

A search using the glitch-robust statistic (i.e.,
Eq.\eqref{eqn_glitch_likelihood}) could be implemented in any number of ways.
Indeed, it could be added to any standard-CW wide-parameter space search.
However, these searches (see, e.g., the recent all-sky searches in LIGO O1 data
\citet{abbott2017allsky, abbott2017allskyeinstein}) already demand massive
computing efforts and adding (at least) two additional search
parameters $\tglitch, \delta f$ would decrease
the sensitivity to standard signals. In this section we investigate
how a glitch-robust search can instead be applied in the follow-up of
potential candidates identified in such
searches. Sec.~\ref{sec_standard} provides some examples of how a
glitching-CW signal may appear in a standard-CW search.

We assume a candidate has been identified by a standard-CW search with some
uncertainty on its phase-evolution parameters $\blambda$. We first discuss the prior
ranges for the phase-evolution and glitch parameters. We then introduce the
necessary tools to quantify the size of a given prior parameter space
before comparing grid- and MCMC-based glitch-robust search methods.

\subsection{Glitch-parameter priors}
\label{sec_priors}

For the standard-CW phase-evolution parameters $\blambda$, the prior (in the absence
of other information) is chosen as uniform over the parameter space of
interest; for a glitch-robust follow-up these will primarily be determined by
the candidate uncertainty. In addition, the glitch-robust search requires priors
on the glitch epochs $\tglitch_\ell$, the magnitude of the
frequency jumps $\delta f_\ell$ and spin-down jumps $\delta\fk_\ell$, and
the number of glitches $\Ng$.

For the number of glitches, a prior could be formed using the glitch rate
observed in the pulsar population. However, dynamically searching over the
number of glitches, which determines the total number of parameters, can be
difficult. For MCMC-based searches, this would require a reversible-jump MCMC
algorithm \citep{gelman2013bayesian}. Instead, we suggest searching over the number of glitches by hand,
namely, perform the search for different numbers of glitches and compare the
results. We will discuss in Sec.~\ref{sec_bayes_factor} how to quantify
this comparison.

For the glitch epochs $\tglitch_\ell$, a uniform prior over the data duration
makes intuitive sense; we also assert that $\tglitch_\ell < \tglitch_{\ell+1}
\forall \ell$. In this work we pragmatically bound $\tglitch_\ell$ between
0.1 and 0.9 of the fractional data duration. This avoids boundary issues where
there is insufficient data to calculate the $\F$-statistic in the first or last
segment and also reduces the parameter space to the region of primary interest,
e.g.,  where a glitch will cause the maximum loss of detection statistic.

Choosing a prior for the jump sizes $\ddl\delta \fk_\ell\ddr$ is more difficult.  Clearly it
should be informed by the glitches seen in the pulsar population and one option
is to use fits to the observed set of glitches in the pulsar population (e.g., see
\citet{ashton2017, fuentes2017}). However, these may be affected by
observational biases. A
simple option is to use a uniform prior on $\{\delta \fk_\ell\}$ between a
minimum and maximum value. For $\delta f^{(0)}_\ell$, one approach is to
set the minimum at zero (excluding anti-glitches where $\delta f^{(0)} < 0$, cf. \citet{archibald2013}) and the maximum at twice the
maximum observed glitches in the pulsar population ($\sim 5\times10^{-5}$~Hz,
see, e.g., \citet{livingstone2009}). Similar approaches can be devised for
higher-order spin-down components.

\subsection{The metric and the size of parameter space}

In setting up any search, it is useful to have a \emph{metric} to understand
distances in the parameter space. Given a detection statistic $d(\theta)$
measured at some set of parameters $\theta$, we first define a mismatch
\begin{align}
\mu(\theta^s, \Delta\theta^s)\equiv
\frac{d(\theta^s) - d(\theta^s + \Delta\theta)}
{d(\theta^s)}
\in [0, 1]\,,
\end{align}
the fractional loss of detection statistic between the exact signal parameters
$\theta^s$ and some other point in the parameter space $\theta^s + \Delta\theta$.

For small mismatches, one may expand and approximate the full mismatch by the
\emph{metric-mismatch}
\begin{align}
\mu(\theta_s, \Delta\theta_s)\approx g_{ij}\Delta\theta^{i}\Delta\theta^{j}
\in [0, \infty)\,,
\end{align}
where $g_{ij}$ is referred to as the \emph{metric} and $\Delta\theta^i$ are
the components of $\Delta\theta$.

As discussed in the next section, the metric is useful in bounding the maximum
loss of detection statistic when setting up grid-based searches. However, one
should note that the metric mismatch is only a good approximation up to
$\mu \gtrsim 0.3-0.5$ \citep{prix2007, wette2013, wette2015}.  Another useful
application of the metric is in calculating $\N^*$, the \emph{approximate
number of unit-mismatch templates} covering the given parameter space
\cite{ashton2018}, which can be understood as a proxy for the \emph{size} of that
parameter space.

Calculation of $\N^*$ requires the ability to calculate the metric. We do not
yet have the metric for the glitch-robust detection statistic defined
in Eq.~\eqref{eqn_glitch_likelihood} (future searches may require this metric in,
for example, a grid-based glitch-robust directed search). Nevertheless, it is
still useful to calculate $\N^*$ using the fully coherent standard detection
statistic over the standard signal parameters, i.e., $\{\fk\}$ and $\sky$. This
can be used a lower bound on the full $\N^*$ for the full parameter space
including the glitch parameters.

\subsection{Grid-based glitch-robust search}

Grid-based (or template-bank) searches compute the detection statistic over a
number of prespecified points in parameter space with the grid of points
covering the prior range. The grid spacing is selected to minimize both the
maximum loss of detection statistic, bounded at some level, and the computing
cost (i.e., to avoid oversampling the space). This spacing is determined using the
metric; for the fully coherent and semicoherent $\F$-statistic see
\citet{wette2013} and \citet{wette2015} respectively. However, as previously discussed, we do not have
the metric for the glitch-robust detection statistic.
So, while we can apply the usual relations to any standard
phase-evolution parameters used in the search and they should approximately
hold, there is no simple way to determine the spacings in $\{\tglitch_\ell\}$ and
$\ddl\delta f^{(k)}\ddr$ that guarantee a bound on the maximum mismatch.

In the absence of the relevant parameter space metric, we will employ a naive method here, simply dividing the full range of each
search parameter into $\Ngrid$ steps. As such, the total number of grid points is
$\Ngrid$ to the power of the number of search dimensions. This choice is not
optimal (as would be the case if one were to derive and use the relevant
metric), but captures many of the salient features of a grid-based search.

\begin{table}[t]
\caption{Relevant simulated signal and noise properties used in
Figs.~\ref{fig_grid_corner} and \ref{fig_MCMC_corner}. $S_n$ is the
noise floor of the detector at the simulated signal frequency.
In the table and figures we use the shorthand~$\fdot\equiv f^{(1)}$.}
\label{tab_sim_params}
\begin{tabular}{llll|lll}
$T$& = & \SI{50}{\day}
& &
$\sqrt{S_n}$& = & \SI{1d-22}{\sqrt\Hz}
\\
$h_0$& = & \num{5d-24}
& &
$\cos\iota$&$=$ & \num{0.5}
\\
$f_s$&$=$ & \SI{30}{\Hz}
& &
$\fdot_s$&$=$ & \SI{-1d-10}{\Hz/\s}
\\
$\delta f_s$&$=$ & \SI{5d-6}{\Hz}
& &
$\tglitch_s$&$=$ & \SI{25}{\day}
\\
$\alpha_s$&$=$ & \SI{83.6292}{\deg}
& &
$\delta_s$&$=$ & \SI{22.0144}{\deg}
\end{tabular}
\end{table}

\begin{figure}[tb]
\centering
\includegraphics[width=0.5\textwidth]{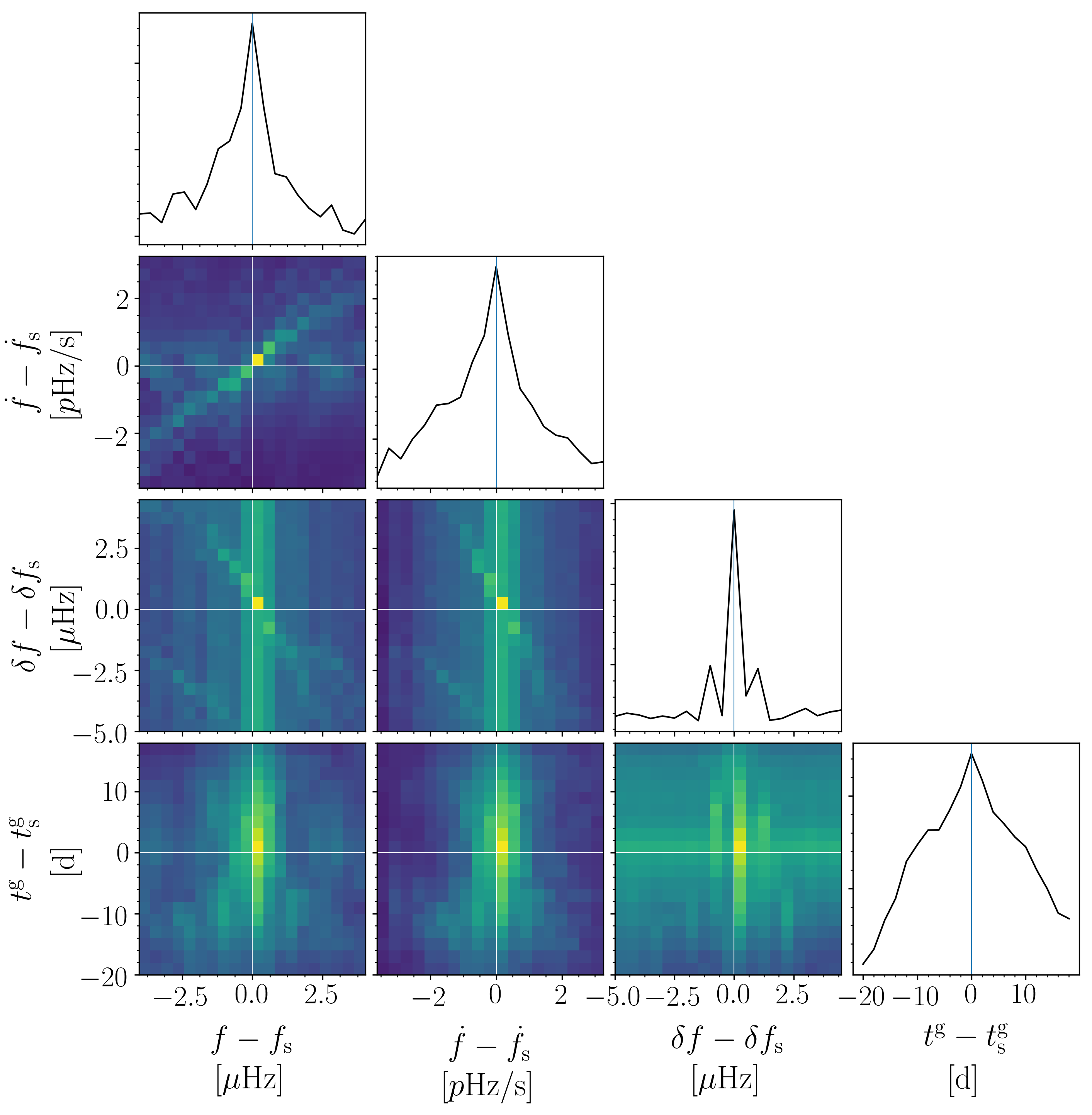}
\caption{Grid-corner plot showing various marginalizations of the glitch-robust
semicoherent $\F$-statistic computed using a grid-based search over the
prior ranges in Table~\ref{tab_unif}. Solid lines
indicate the simulated signal parameters.}
\label{fig_grid_corner}
\end{figure}

As an example of the grid-based method, we simulate a glitching signal in
Gaussian noise with the properties given in Table~\ref{tab_sim_params}.  Note
that the glitch occurs in frequency alone, i.e., $\delta \fk=0$ for $k>0$.
We then perform a grid-based search  over $\{f, \fdot, \tglitch, \delta
f\}$ with $\Ngrid=20$; in this search the sky-location, $\sky$, is fixed to
that of the simulated value. The prior ranges are given in
Table~\ref{tab_unif}. In Fig.~\ref{fig_grid_corner}, we plot the
semicoherent glitch-robust $\F$-statistic in a \emph{grid-corner} plot. This,
as with the corner plots used in MCMC parameter estimation, displays
the marginalized detection statistic for all one- and two-dimensional
combinations.

\begin{table}[th]
\caption{Priors used for the search parameters;
the subscript s indicates the simulation values given in
Table~\ref{tab_sim_params} and $\tglitch$ is defined from the start of the
observation span.
For the uncertainty in $f$ and $\fdot$, the number of fully coherent unit-mismatch templates is $\N^*=1000$.}
\label{tab_unif}
\begin{tabular}{lll}
& & Uniform prior range\\\hline
$f$&$\sim$ & $f_s \pm 4.0\times10^{-6}$ Hz\\
$\fdot$&$\sim$ & $\fdot_s \pm 1.8\times10^{-12}$ Hz/s\\
$\delta f$&$\sim$ & $[0, 5\times10^{-5}]$ Hz\\
$\tglitch $&$\sim$ & $[5, 45]$ days\\
\end{tabular}
\end{table}

The grid spacing in this instance is sufficiently fine to provide reasonably
good parameter estimation. For detection purposes it may even suffice to have sparser
template coverage in the glitch time (where the signal appears quite wide
compared to the prior range). At a fixed computing cost, this would allow for denser
coverage in other parameters where the signal is narrower compared to the
prior range.

\subsection{MCMC-based glitch-robust search}

\begin{figure}[tb]
\centering
\includegraphics[width=0.5\textwidth]{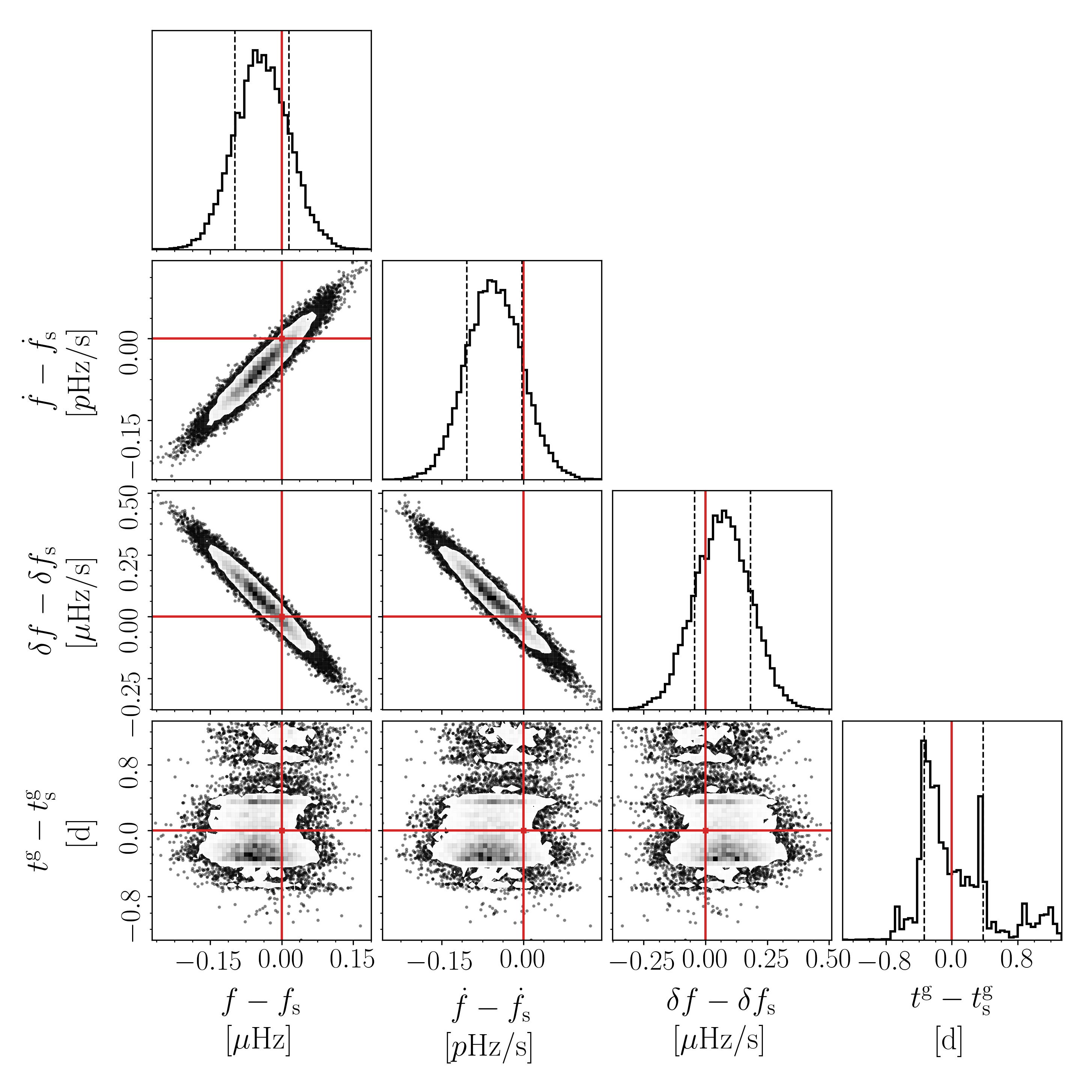}
\caption{Corner plot showing various marginalizations of the exponential of
the glitch-robust semicoherent $\F$-statistic computed using an MCMC-based
search. Solid lines indicate the simulated signal parameters while dashed
lines (on the one-dimensional histograms) indicate 1-$\sigma$ quantiles.
Figure generated using the \textsc{corner} \citep{corner} package.}
\label{fig_MCMC_corner}
\end{figure}

MCMC-based standard-CW searches have already been used with success
\citep{christensendupuis2004, veitch2007, umstatter2004, abbott2010}.  Recently
we demonstrated \citep{ashton2018} that this success relies on the size of
the parameter space being sufficiently small, as quantified by $\N^*$.
Namely it was found that typically $\N^* \lesssim 1000$ is a good guideline,
but this can depend on the exact MCMC setup. For too-large parameter spaces, the MCMC algorithm
tends to fail to converge to the signal peak in a reasonable amount of time.

For the follow-up of candidates from wide-parameter space searches, the size of
the phase-evolution parameter space (i.e., the candidate uncertainty) is well
constrained (or, this can be ensured by performing a refinement step). It is
not possible to calculate $\N^*$ for a glitch-robust detection statistic without
the metric. However, in practice, we find that for typical glitch size and
rates seen in the pulsar population \citep{ashton2018, fuentes2017}
and typical observing spans, an MCMC-based
glitch-robust search is effective at converging on simulated signals. For
longer observing spans (or if allowing for larger glitches than those observed
in the pulsar population) further work will need to be carried out to ensure
the method is robust.

The advantage of an MCMC-based, instead of a grid-based, approach is that there
is no requirement to predetermine the grid points. In effect, the ensemble MCMC
sampler adapts to the topology of the maxima during the burn-in phase (for a
more detailed overview of MCMC-based CW search methods see \citet{ashton2018}).

To illustrate the results of an MCMC search, we run it on the same data set
used to produce Fig.~\ref{fig_grid_corner} (simulation properties are given
in Table~\ref{tab_sim_params}) with the same uniform priors, as
given in Table~\ref{tab_unif}. In Fig.~\ref{fig_MCMC_corner} we plot
the resulting corner plot.

MCMC searches produce samples from the posterior, which, if the signal is
successfully identified, usually occupies only a small fraction of the prior
range. As a result, an MCMC search does not produce a posterior over the whole
prior range, but only over the region of interest. As a consequence of this the
range shown in Fig.~\ref{fig_grid_corner} is much larger than that of
Fig.~\ref{fig_MCMC_corner}: the latter shows the range of the posterior peak
only while the former shows the entire prior range. Moreover, we note that in
Fig.~\ref{fig_grid_corner} for the grid-based search we plot the
$\F$-statistic, corresponding to the (marginalized) log-likelihood ratio.  On
the other hand, in Fig.~\ref{fig_MCMC_corner} for the MCMC-based search we plot
the estimated posterior which in this instance, where we use uniform priors, is
proportional to the likelihood, corresponds to the exponential of the
$\F$-statistic. This is why the peak looks much narrower compared to
Fig.~\ref{fig_grid_corner} while showing in principle the same
likelihood-function.

\subsection{Comparing grid- and MCMC-based searches}

In order to provide a simple comparison between grid- and MCMC-based searches, we run a
Monte-Carlo study. We produce 500 data sets containing a
simulated signal with a single glitch in Gaussian noise. Such a signal,
perfectly matched, has a predicted $\twoFtilde$ of approximately~$330$. The noise, amplitude
and standard phase evolution parameters are given in
Table~\ref{tab_sim_params}, except that we jitter the frequency and spin-down,
picking their value uniformly from within the inner half of the prior region
given in Table~\ref{tab_unif}. We also select the glitch epoch from the
distribution given in this table. Meanwhile, for the glitch magnitude, we
sample from the observed pulsar population distribution \citep{ashton2017};
while the aim of the section is to compare search methods, this choice of
simulation distribution allows us to also verify that the naive priors
are robust to a more astrophysically motivated population distribution.

\begin{figure}[tb]
\centering
\includegraphics[width=0.5\textwidth]{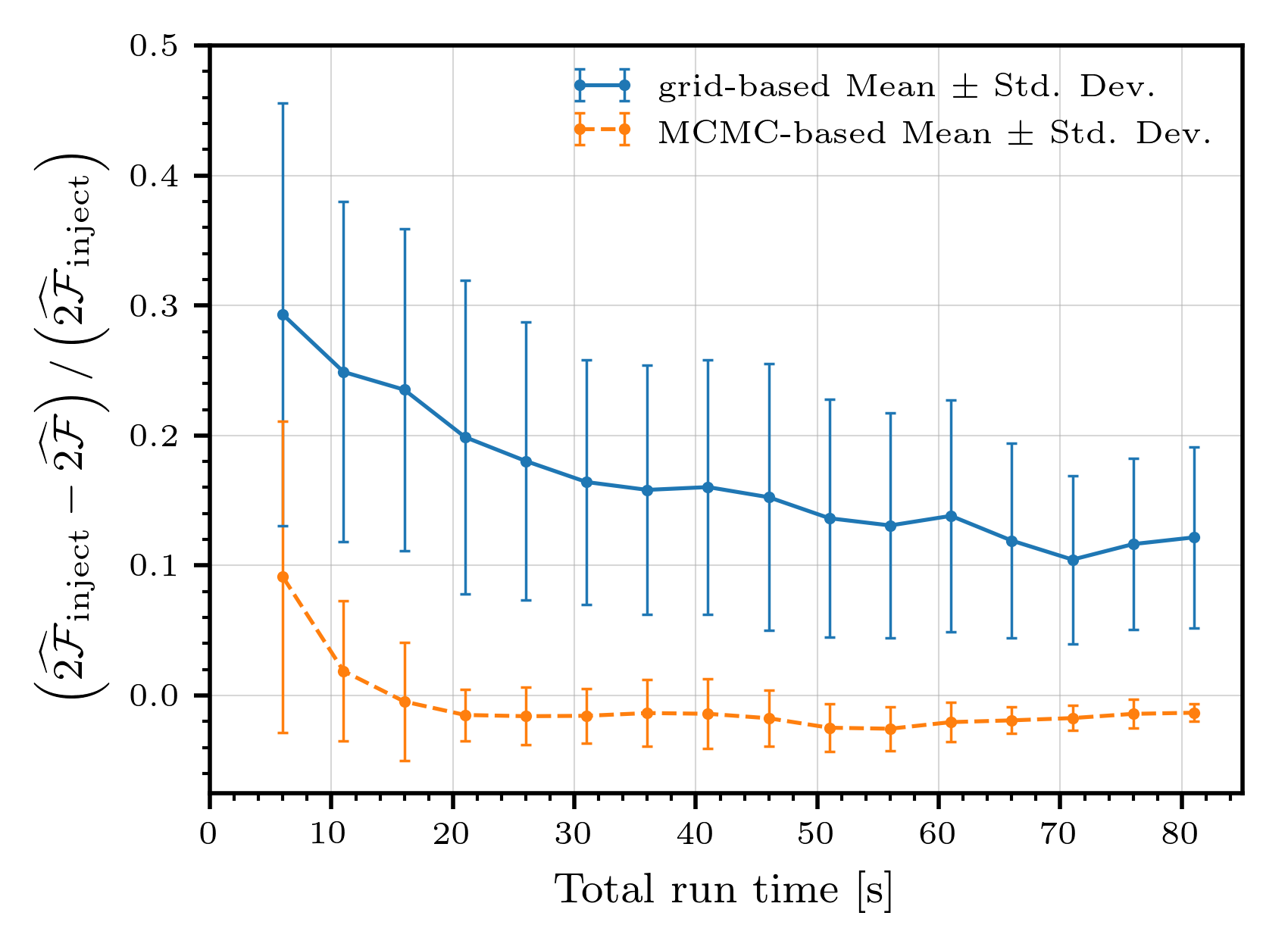}
\caption{Comparison of the relative maximum $\twoFhat$ found by each method
compared to $\twoFhat_\inject$, the value calculated at the
simulated signal parameters $\blambda_s$. All timings performed on an
Intel® Core™ i7-7820HQ CPU @ 2.90GHz processor.}
\label{fig_compare}
\end{figure}

Varying the required computation time (for the grid-based search, by varying
the number of grid points; for the MCMC-based search, by varying the number of
steps taken), in Fig.~\ref{fig_compare} we plot the relative difference between the
recovered maximum $\twoFhat$ (for each method) and $\twoFhat_\inject = \twoFhat(\blambda_s)$, the
statistic measured at the simulated signal parameters $\blambda_s$. Due to the
presence of noise, the actual maximum $\twoFhat$ will typically not occur at the signal parameters $\blambda_s$
but be slightly offset and we therefore generally expect the maximum recovered $\twoFhat > \twoFhat_\inject$, provided the search method
manages to localize the maximum $\twoFhat$ well enough.

From this figure, it is evident that at the same run-time, the MCMC-based
search outperforms the grid-based search, with the majority of points finding a
larger detection statistic than $\twoFhat_\inject$. This is to be
expected since the MCMC search is operating optimally (i.e., the size of
parameter space is sufficiently small). As such, the MCMC quickly converges
to the maximum while a grid-based search spends most of the computing time
calculating the detection statistic for points not near to the signal peak.
For added context, Figs~\ref{fig_grid_corner} and \ref{fig_MCMC_corner} both
have an approximate run-time of 90s; for the grid-based search, the peak is
only sampled a handful of times yet almost all of the MCMC samples (by design)
come from the peak.

\section{Glitching vs.\ standard-CW Bayes factor}
\label{sec_bayes_factor}

We now discuss how to quantify whether a signal is glitching and how many
glitches best explain the data. We do this using a \emph{Bayes factor}, the
ratio of likelihoods for data $\data$ under two hypotheses. If $\Hn$ implies
that the data contains only Gaussian noise while $\Hs$ implies that it contains an
CW signal in addition to noise, then
\begin{align}
\Bsn(\data) \equiv \frac{P(\data| \Hs)}{P(\data| \Hn)}\,.
\end{align}

It can be shown \citep{prix2009, prix2011,
ashton2018} that the signal vs.\ noise Bayes factor at fixed phase-evolution parameters $\blambda$ is
\begin{equation}
\Bsn(\data; \blambda, \Nseg) =
 \left(\frac{70}{\rhohatmax^4}\right)^{\Nseg}
e^{\hat{\F}(\data; \blambda, \Nseg)}\,,
\end{equation}
where $\hat{\F}(\data; \blambda, \Nseg)$ is the $N$-segment semicoherent
$\F$-statistic defined in Eq.~\eqref{eqn_semi} and $\rhohatmax$
is an arbitrary upper cutoff on the prior range in signal strength \citep{prix2011}.

Similarly, defining $\Hg$ as the glitching-signal hypothesis, we see that the
targeted (in the sense that it depends on the model parameter) glitching-signal
vs.\ noise Bayes factor is
\begin{align}
\begin{split}
\Bgn\left(\data; \blambda, \ddl\delta \fk_\ell\ddr, \{\tglitch_\ell\}, \Ng\right) & \\
& \hspace{-25mm} \equiv
\left(\frac{70}{\rhohatmax^4}\right)^{\Ng+1}
e^{\hat{\F}\left(\data; \blambda, \ddl\delta \fk_\ell\ddr, \{\tglitch_\ell\}\right)}\,,
\end{split}
\end{align}
where the exponent is the glitch-robust semicoherent $\F$-statistic,
defined in Eq.~\eqref{eqn_glitch_likelihood}.

After marginalizing the targeted Bayes factor we get the signal vs.\ noise Bayes factor; i.e., for
the standard search,
\begin{equation}
\Bsn(\data; \Nseg) =
\int \Bsn(\data; \blambda, \Nseg) P(\blambda| \Hs) d\blambda,
\end{equation}
while for the semicoherent glitch-robust
\begin{align}
\begin{split}
\Bgn(\data; \Ng) & = \int
\Bsn\left(\data; \blambda, \ddl\delta \fk_\ell\ddr, \{\tglitch_\ell\}, \Ng\right) \times \\
& \hspace{9mm}
P\left(\blambda, \ddl\delta \fk_\ell\ddr, \{\tglitch_\ell\}| \Hs\right)\\
& \hspace{9mm}
d\blambda d\ddl\delta \fk_\ell\ddr d\{\tglitch_\ell\}\,.
\end{split}
\end{align}

The arbitrary prior cutoff $\rhohatmax$ makes it difficult to interpret either
of these Bayes factors by themselves: one can tune the Bayes factor by
arbitrary changes in the prior. However, if we define
\begin{align}
\Bgs(\data, \Ng) \equiv \frac{\Bgn(\data, \Ng)}{\Bsn(\data, \Nseg=\Ng+1)}\,,
\end{align}
the glitching-CW vs.\ standard-CW Bayes factor, then the arbitrary prior
cutoff cancels and we are left with an interpretable Bayes factor for whether the
signal is glitching or not.

Calculation of the Bayes factor can be done by either a grid-based (using
numerical integration of a dense sampling of the posterior) or MCMC-based
method (using thermodynamic integration \citep{goggans2004}). In future,
we intend to extend the functionality to include nested sampling
\citep{skilling2006nested} which will improve the robustness of the
evidence calculation (see, e.g., Ref.~\citep{allison2014} for a comparison).

\begin{figure}[tb]
\centering
\includegraphics[width=0.5\textwidth]{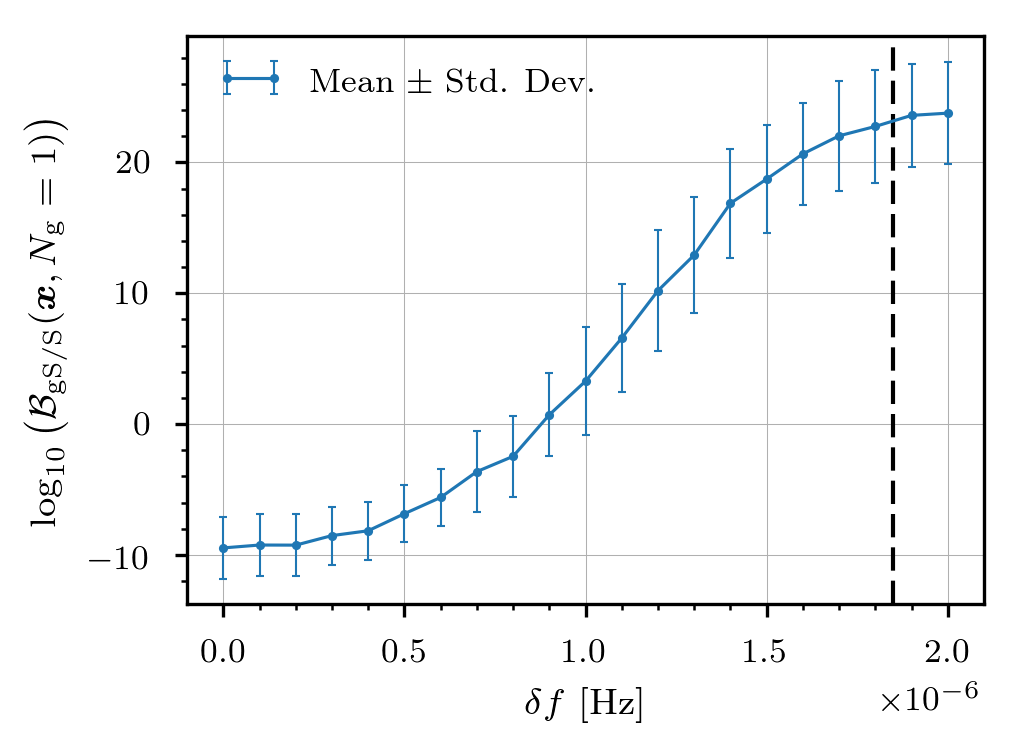}
\caption{Monte Carlo study of the $\Bgs(\data, \Ng=1)$ Bayes factor as a
function of the simulated glitch magnitude. A dashed vertical line indicates
the value of Eq.~\eqref{eqn_deltaF_unit} for the 50-day duration used in this
study.}
\label{fig_bayes}
\end{figure}

To understand the behaviour of $\Bgs(\data, \Ng)$ as a function of the glitch
magnitude, we run a Monte Carlo study, simulating 100 data sets (for
each $\delta f$)
with a glitching signal in Gaussian noise. We use the parameters given in
Table~\ref{tab_sim_params}, except $\delta f$ which we vary systematically over
a relevant domain.  For each data set, we run a glitch-robust semicoherent
MCMC search with $\Ng=1$ along with a semicoherent MCMC search with $\Nseg=2$
and calculate the resulting Bayes factor. The MCMC parameters are chosen
such that the log Bayes factors are estimated to within a few percent. In
Fig.~\ref{fig_bayes} we plot the mean and standard deviation
calculated over all data sets. We see that for small glitches, the Bayes factor
prefers the standard signal hypothesis.  But, once glitches are sufficiently
large, the glitching-signal hypothesis is preferred.

To determine the preferred number of glitches, $\Bgs(\data, \Ng)$ can
be calculated for different $\Ng$ and interpreted as a posterior over $\Ng$.
Large numbers of glitches, $\gtrsim 10$ say, may be difficult to handle and require
some tuning of the MCMC sampler.

Fig.~\ref{fig_bayes} illustrates that the glitch-robust search (as a function
of the glitch size) plateaus above a certain minimum glitch size.  An
approximate way to characterize this size is to use the averaged (over
glitch-time) single-glitch metric mismatch expressions derived in
\citet{ashton2017}. Note that this is the metric for a standard CW search of a
glitching signal and not the metric for the glitch-robust statistic introduced
in Sec~\ref{sec_glitch_robust}.  For example, for a fully coherent search at a
fixed sky-location over frequency and spin-down, the minimum average metric
mismatch is given by $\tilde{\mu}=(\pi T \delta f)^2/630$.  Setting the
metric-mismatch to unity and inverting gives
\begin{align}
\delta f = \frac{\sqrt{630}}{\pi T}\,,
\label{eqn_deltaF_unit}
\end{align}
a rough order-of-magnitude estimate of the glitch size for which a standard-CW
search would be sufficiently affected by the glitch that the glitch hypothesis
will be preferred. Similar results can be derived for jumps in higher-order frequency spin-downs
using the corresponding components of the glitch metric.

In Fig.~\ref{fig_bayes}, we plot the value of Eq.~\eqref{eqn_deltaF_unit},
given the 50-day duration of data used.  Notably, this agrees with the point at
which the Bayes factor begins to plateau.

\section{Identifying glitching signal in standard searches}
\label{sec_standard}

In order to identify when a signal candidate from a standard-CW semicoherent search
might best be followed up using a glitch-robust method, we now
discuss the behaviour of glitching signals in a standard-CW search.

\subsection{Multiple modes}

One indicator of a glitching signal in a standard-CW search is the existence of
multiple peaks in the detection statistic resulting from the template matching
different parts of the signals. How exactly this behaviour manifests depends on
the magnitude and size of the glitches, the data span, and the search setup.

Considering a signal which undergoes a single glitch with a jump $\{\delta
\fk\}$, we can identify two limiting cases depending on whether the glitch size
is smaller or larger than a critical glitch size $\delta \fk_c$ : if $\delta
\fk \ll \delta \fk_c $, the effect of the glitch is negligible within the
search setup; if instead, $\delta\fk \gg \delta\fk_c$, the signal can be
thought of as two transient CWs, and we will find two distinct peaks in the
detection statistic corresponding to the pre- and post-glitch signal
parameters. Between these two extremes, when $\delta\fk \sim \delta \fk_c$, the
resulting structure in the detection statistic can be quite complicated. If
required, the critical glitch size can be estimated from the single-glitch
metric mismatches derived in \citet{ashton2017}.

\begin{figure}[tb]
\centering
\includegraphics[width=0.5\textwidth]{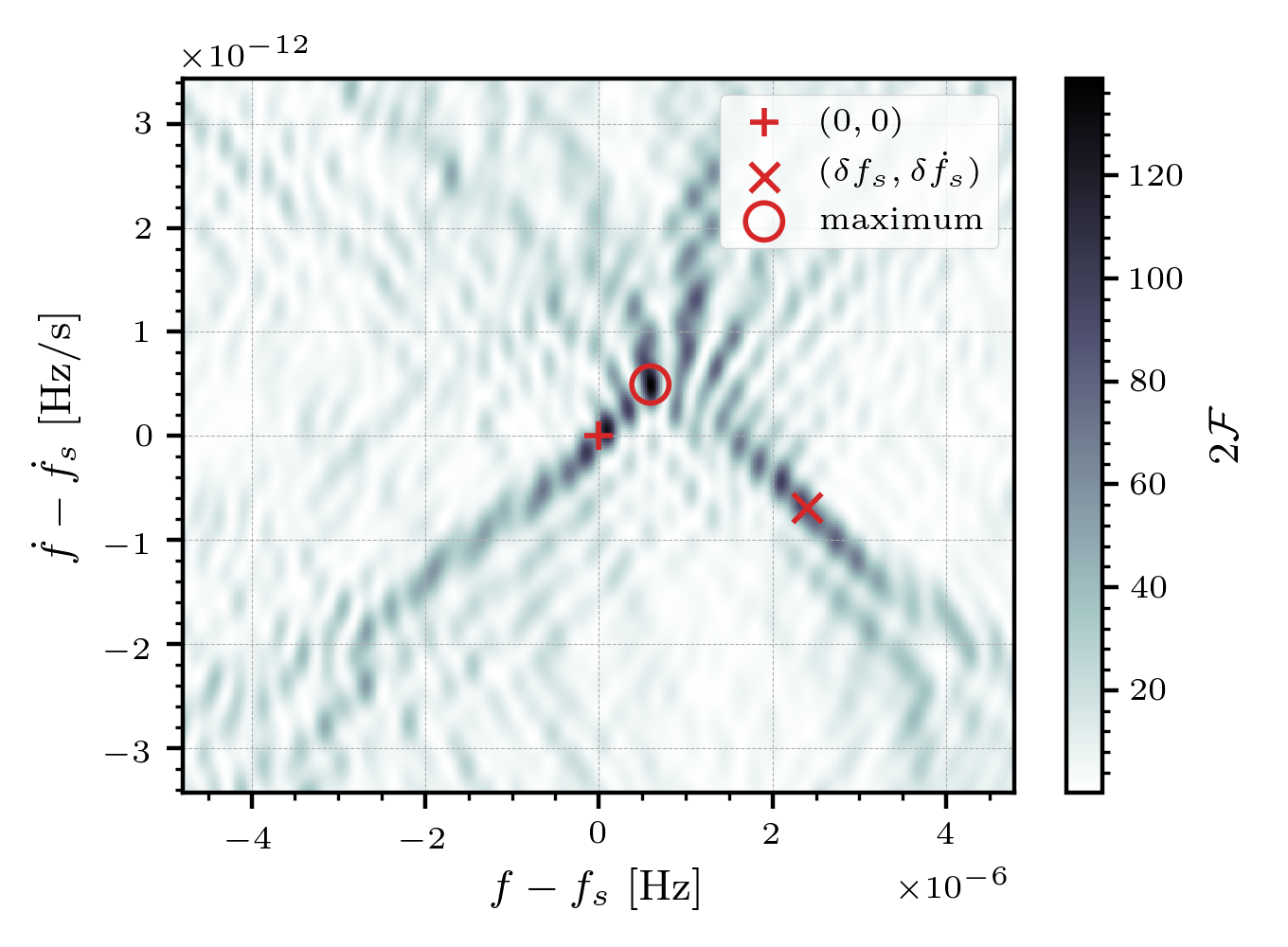}
\caption{The fully coherent $\twoFtilde$ computed over a grid in $f$ and
$\dot{f}$ for the simulated glitching signal.}
\label{fig_glitch_grid}
\end{figure}

In order to illustrate this intermediate case, in
Fig.~\ref{fig_glitch_grid} we show a standard-CW fully coherent grid search
over frequency and spin-down for a simulated glitching signal. The
simulation properties are given in Table~\ref{tab_sim_params}, except that we
set $\delta f =\SI{2.4e-6}{Hz}$ and $\delta\fdot=\SI{-6.9d-13}{Hz/s}$.
As the reference time and glitch time coincide, for the fully coherent search,
the pre-glitch frequency and derivative is $f_s$, $\fdot_s$ while the
post-glitch frequency and derivative are $f_s+\delta f$, $\fdot_s +
\delta\fdot$. Two distinct ellipsoid patterns can be observed centered on the
locations of the pre and post-glitch signals, but the maximum does not coincide
with either.

Having multiple peaks in the frequency and its derivatives might be
expected, but we typically also find multiple peaks in the sky position,
even though the sky position of the source does not vary over a glitch.
This is because by allowing the sky position to vary, the standard template fit
to the glitching signal can be improved; this can happen in multiple ways,
resulting in multiple peaks and will in general result in biases in the
recovered sky position.

\subsection{Sliding windows}

\begin{figure}[tb]
\centering
\includegraphics[width=0.5\textwidth]{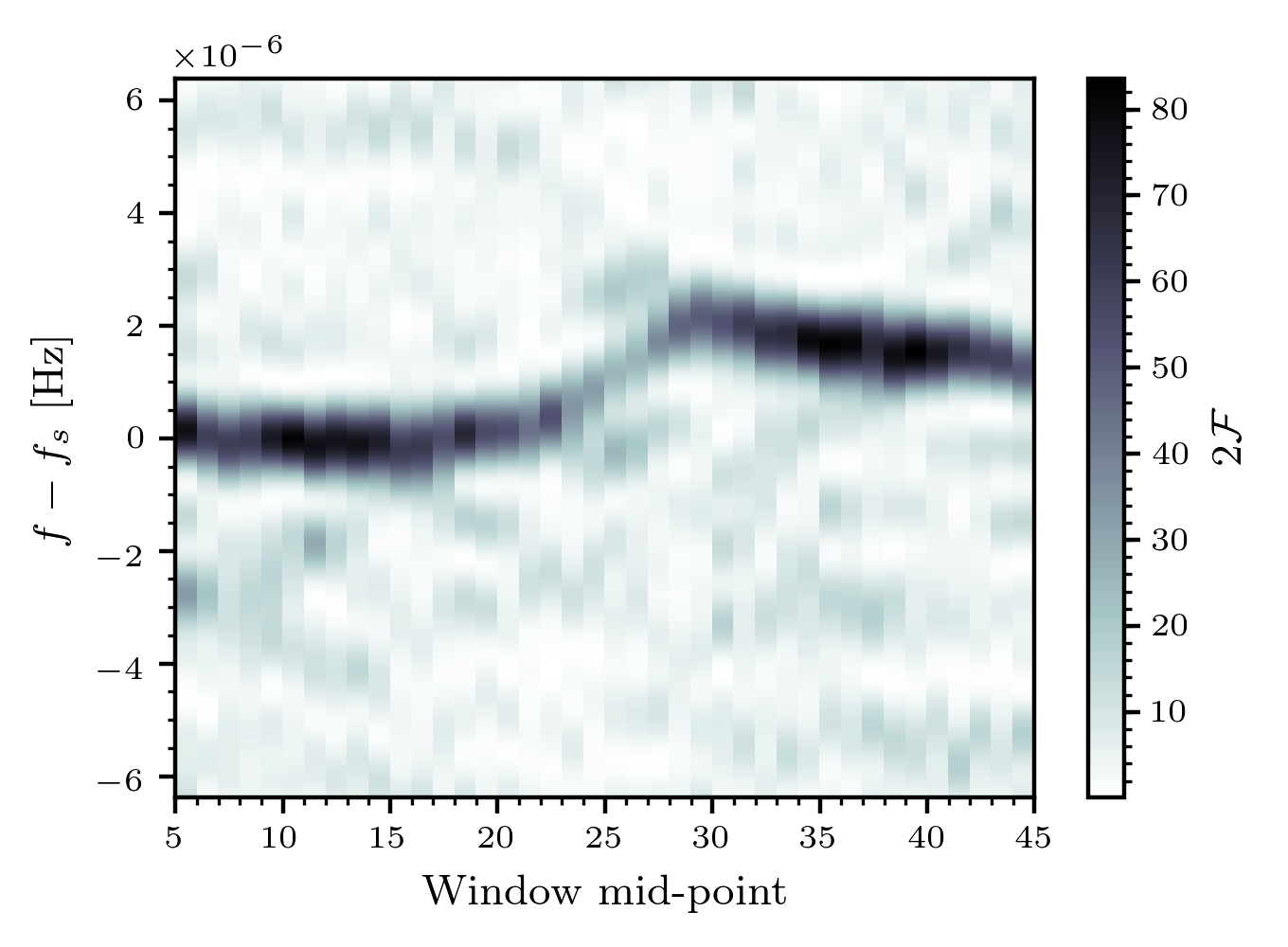}
\caption{Illustration of the frequency sliding-window, a useful diagnostic tool
for identifying glitching signals. In this example, a 10-day window is slid in
increments of 1 day.}
\label{fig_sliding_window}
\end{figure}

A sliding window can be another simple, but powerful diagnostic test for a
glitching signal. Fixing all other values to those of the maximum posterior
estimate (or a set of parameters sufficiently close to the peak), the detection
statistic is computed for a range of frequencies in an overlapping sliding
time-window over the total data span. One could also do this for the frequency
derivative (or any other parameter). Stacking the results together into a
colour plot, if the signal is sufficiently strong, the glitch can easily be
discerned from the change in frequency. We provide an example in
Fig.~\ref{fig_sliding_window} using the same data set used to produce
Fig.~\ref{fig_glitch_grid}.

\section{Discussion}
\label{sec_discussion}

We have described a semicoherent glitch-robust detection statistic for use in
evaluating if candidates found in wide-parameter space searches are glitching
signals. This simple method adapts standard search routines, using a signal
model that includes glitches as a set of instantaneous changes in the frequency
and higher-order spin-downs at a set of glitch epochs.

Comparing grid- and MCMC-based search methods, we find that the MCMC-based
search is a superior method for performing glitch-robust searches of candidates
from wide-parameter space searches. For the same
computing cost it is able to better identify the maximum and perform parameter
estimation vital to interpretation. MCMC-based glitch-robust searches are, for a
suitable candidate uncertainty level, computationally cheap to run and provide
parameter estimation and evidence estimates. Moreover, an MCMC-based method
does not require a pre-specified grid template. We therefore
recommend that such glitch-robust MCMC-based methods be used in the follow-up
of candidates identified in wide-parameter searches.

The methods introduced in this paper have been implemented in the
package \textsc{pyfstat} \citep{pyfstat}. Source code along
with all examples in this work can be found at
\url{https://gitlab.aei.uni-hannover.de/GregAshton/PyFstat}.

\section{Acknowledgements}

The authors kindly thank the members of the Continuous Waves group of the LIGO
Scientific Collaboration and the Virgo Scientific Collaboration for useful
feedback and discussion during the development of this work. We thank Sylvia
Zhu for cogent and useful comments during the preparation of the manuscript.
DIJ acknowledges funding from STFC through grant number ST/M000931/1. This
article has been assigned the document number LIGO-P1800057.

\def\mnras{Mon. Notices Royal Astron. Soc.}
\def\cqg{Class. Quantum Grav.}
\bibliography{bibliography}

\begin{thebibliography}{29}%
\makeatletter
\providecommand \@ifxundefined [1]{%
 \@ifx{#1\undefined}
}%
\providecommand \@ifnum [1]{%
 \ifnum #1\expandafter \@firstoftwo
 \else \expandafter \@secondoftwo
 \fi
}%
\providecommand \@ifx [1]{%
 \ifx #1\expandafter \@firstoftwo
 \else \expandafter \@secondoftwo
 \fi
}%
\providecommand \natexlab [1]{#1}%
\providecommand \enquote  [1]{``#1''}%
\providecommand \bibnamefont  [1]{#1}%
\providecommand \bibfnamefont [1]{#1}%
\providecommand \citenamefont [1]{#1}%
\providecommand \href@noop [0]{\@secondoftwo}%
\providecommand \href [0]{\begingroup \@sanitize@url \@href}%
\providecommand \@href[1]{\@@startlink{#1}\@@href}%
\providecommand \@@href[1]{\endgroup#1\@@endlink}%
\providecommand \@sanitize@url [0]{\catcode `\\12\catcode `\$12\catcode
  `\&12\catcode `\#12\catcode `\^12\catcode `\_12\catcode `\%12\relax}%
\providecommand \@@startlink[1]{}%
\providecommand \@@endlink[0]{}%
\providecommand \url  [0]{\begingroup\@sanitize@url \@url }%
\providecommand \@url [1]{\endgroup\@href {#1}{\urlprefix }}%
\providecommand \urlprefix  [0]{URL }%
\providecommand \Eprint [0]{\href }%
\providecommand \doibase [0]{http://dx.doi.org/}%
\providecommand \selectlanguage [0]{\@gobble}%
\providecommand \bibinfo  [0]{\@secondoftwo}%
\providecommand \bibfield  [0]{\@secondoftwo}%
\providecommand \translation [1]{[#1]}%
\providecommand \BibitemOpen [0]{}%
\providecommand \bibitemStop [0]{}%
\providecommand \bibitemNoStop [0]{.\EOS\space}%
\providecommand \EOS [0]{\spacefactor3000\relax}%
\providecommand \BibitemShut  [1]{\csname bibitem#1\endcsname}%
\let\auto@bib@innerbib\@empty
\bibitem [{\citenamefont {{Abbott}}\ \emph
  {et~al.}(2017{\natexlab{a}})\citenamefont {{Abbott}}, \citenamefont
  {{Abbott}}, \citenamefont {{Abbott}}, \citenamefont {{Acernese}},
  \citenamefont {{Ackley}}, \citenamefont {{Adams}}, \citenamefont {{Adams}},
  \citenamefont {{Addesso}}, \citenamefont {{Adhikari}}, \citenamefont
  {{Adya}},\ and\ \citenamefont {et~al.}}]{abbott2017allsky}%
  \BibitemOpen
  \bibfield  {author} {\bibinfo {author} {\bibfnamefont {B.~P.}\ \bibnamefont
  {{Abbott}}}, \bibinfo {author} {\bibfnamefont {R.}~\bibnamefont {{Abbott}}},
  \bibinfo {author} {\bibfnamefont {T.~D.}\ \bibnamefont {{Abbott}}}, \bibinfo
  {author} {\bibfnamefont {F.}~\bibnamefont {{Acernese}}}, \bibinfo {author}
  {\bibfnamefont {K.}~\bibnamefont {{Ackley}}}, \bibinfo {author}
  {\bibfnamefont {C.}~\bibnamefont {{Adams}}}, \bibinfo {author} {\bibfnamefont
  {T.}~\bibnamefont {{Adams}}}, \bibinfo {author} {\bibfnamefont
  {P.}~\bibnamefont {{Addesso}}}, \bibinfo {author} {\bibfnamefont {R.~X.}\
  \bibnamefont {{Adhikari}}}, \bibinfo {author} {\bibfnamefont {V.~B.}\
  \bibnamefont {{Adya}}}, \ and\ \bibinfo {author} {\bibnamefont {et~al.}},\
  }\bibfield  {title} {\enquote {\bibinfo {title} {{All-sky search for periodic
  gravitational waves in the O1 LIGO data}},}\ }\href {\doibase
  10.1103/PhysRevD.96.062002} {\bibfield  {journal} {\bibinfo  {journal}
  {\prd}\ }\textbf {\bibinfo {volume} {96}},\ \bibinfo {eid} {062002} (\bibinfo
  {year} {2017}{\natexlab{a}})},\ \Eprint {http://arxiv.org/abs/1707.02667}
  {arXiv:1707.02667 [gr-qc]} \BibitemShut {NoStop}%
\bibitem [{\citenamefont {{Abbott}}\ \emph
  {et~al.}(2017{\natexlab{b}})\citenamefont {{Abbott}}, \citenamefont
  {{Abbott}}, \citenamefont {{Abbott}}, \citenamefont {{Acernese}},
  \citenamefont {{Ackley}}, \citenamefont {{Adams}}, \citenamefont {{Adams}},
  \citenamefont {{Addesso}}, \citenamefont {{Adhikari}}, \citenamefont
  {{Adya}},\ and\ \citenamefont {et~al.}}]{abbott2017allskyeinstein}%
  \BibitemOpen
  \bibfield  {author} {\bibinfo {author} {\bibfnamefont {B.~P.}\ \bibnamefont
  {{Abbott}}}, \bibinfo {author} {\bibfnamefont {R.}~\bibnamefont {{Abbott}}},
  \bibinfo {author} {\bibfnamefont {T.~D.}\ \bibnamefont {{Abbott}}}, \bibinfo
  {author} {\bibfnamefont {F.}~\bibnamefont {{Acernese}}}, \bibinfo {author}
  {\bibfnamefont {K.}~\bibnamefont {{Ackley}}}, \bibinfo {author}
  {\bibfnamefont {C.}~\bibnamefont {{Adams}}}, \bibinfo {author} {\bibfnamefont
  {T.}~\bibnamefont {{Adams}}}, \bibinfo {author} {\bibfnamefont
  {P.}~\bibnamefont {{Addesso}}}, \bibinfo {author} {\bibfnamefont {R.~X.}\
  \bibnamefont {{Adhikari}}}, \bibinfo {author} {\bibfnamefont {V.~B.}\
  \bibnamefont {{Adya}}}, \ and\ \bibinfo {author} {\bibnamefont {et~al.}},\
  }\bibfield  {title} {\enquote {\bibinfo {title} {{First low-frequency
  Einstein@Home all-sky search for continuous gravitational waves in Advanced
  LIGO data}},}\ }\href {\doibase 10.1103/PhysRevD.96.122004} {\bibfield
  {journal} {\bibinfo  {journal} {\prd}\ }\textbf {\bibinfo {volume} {96}},\
  \bibinfo {eid} {122004} (\bibinfo {year} {2017}{\natexlab{b}})},\ \Eprint
  {http://arxiv.org/abs/1707.02669} {arXiv:1707.02669 [gr-qc]} \BibitemShut
  {NoStop}%
\bibitem [{\citenamefont {{Hobbs}}\ \emph {et~al.}(2010)\citenamefont
  {{Hobbs}}, \citenamefont {{Lyne}},\ and\ \citenamefont
  {{Kramer}}}]{hobbs2010}%
  \BibitemOpen
  \bibfield  {author} {\bibinfo {author} {\bibfnamefont {G.}~\bibnamefont
  {{Hobbs}}}, \bibinfo {author} {\bibfnamefont {A.~G.}\ \bibnamefont {{Lyne}}},
  \ and\ \bibinfo {author} {\bibfnamefont {M.}~\bibnamefont {{Kramer}}},\
  }\bibfield  {title} {\enquote {\bibinfo {title} {{An analysis of the timing
  irregularities for 366 pulsars}},}\ }\href {\doibase
  10.1111/j.1365-2966.2009.15938.x} {\bibfield  {journal} {\bibinfo  {journal}
  {\mnras}\ }\textbf {\bibinfo {volume} {402}},\ \bibinfo {pages} {1027--1048}
  (\bibinfo {year} {2010})},\ \Eprint {http://arxiv.org/abs/0912.4537}
  {arXiv:0912.4537} \BibitemShut {NoStop}%
\bibitem [{\citenamefont {{Espinoza}}\ \emph {et~al.}(2011)\citenamefont
  {{Espinoza}}, \citenamefont {{Lyne}}, \citenamefont {{Stappers}},\ and\
  \citenamefont {{Kramer}}}]{espinoza2011}%
  \BibitemOpen
  \bibfield  {author} {\bibinfo {author} {\bibfnamefont {C.~M.}\ \bibnamefont
  {{Espinoza}}}, \bibinfo {author} {\bibfnamefont {A.~G.}\ \bibnamefont
  {{Lyne}}}, \bibinfo {author} {\bibfnamefont {B.~W.}\ \bibnamefont
  {{Stappers}}}, \ and\ \bibinfo {author} {\bibfnamefont {M.}~\bibnamefont
  {{Kramer}}},\ }\bibfield  {title} {\enquote {\bibinfo {title} {{A study of
  315 glitches in the rotation of 102 pulsars}},}\ }\href {\doibase
  10.1111/j.1365-2966.2011.18503.x} {\bibfield  {journal} {\bibinfo  {journal}
  {\mnras}\ }\textbf {\bibinfo {volume} {414}},\ \bibinfo {pages} {1679--1704}
  (\bibinfo {year} {2011})},\ \Eprint {http://arxiv.org/abs/1102.1743}
  {arXiv:1102.1743 [astro-ph.HE]} \BibitemShut {NoStop}%
\bibitem [{\citenamefont {{Fuentes}}\ \emph {et~al.}(2017)\citenamefont
  {{Fuentes}}, \citenamefont {{Espinoza}}, \citenamefont {{Reisenegger}},
  \citenamefont {{Shaw}}, \citenamefont {{Stappers}},\ and\ \citenamefont
  {{Lyne}}}]{fuentes2017}%
  \BibitemOpen
  \bibfield  {author} {\bibinfo {author} {\bibfnamefont {J.~R.}\ \bibnamefont
  {{Fuentes}}}, \bibinfo {author} {\bibfnamefont {C.~M.}\ \bibnamefont
  {{Espinoza}}}, \bibinfo {author} {\bibfnamefont {A.}~\bibnamefont
  {{Reisenegger}}}, \bibinfo {author} {\bibfnamefont {B.}~\bibnamefont
  {{Shaw}}}, \bibinfo {author} {\bibfnamefont {B.~W.}\ \bibnamefont
  {{Stappers}}}, \ and\ \bibinfo {author} {\bibfnamefont {A.~G.}\ \bibnamefont
  {{Lyne}}},\ }\bibfield  {title} {\enquote {\bibinfo {title} {{The glitch
  activity of neutron stars}},}\ }\href {\doibase 10.1051/0004-6361/201731519}
  {\bibfield  {journal} {\bibinfo  {journal} {Astron. Astrophys.}\ }\textbf
  {\bibinfo {volume} {608}},\ \bibinfo {eid} {A131} (\bibinfo {year} {2017})},\
  \Eprint {http://arxiv.org/abs/1710.00952} {arXiv:1710.00952 [astro-ph.HE]}
  \BibitemShut {NoStop}%
\bibitem [{\citenamefont {{Jones}}(2004)}]{jones2004}%
  \BibitemOpen
  \bibfield  {author} {\bibinfo {author} {\bibfnamefont {D.~I.}\ \bibnamefont
  {{Jones}}},\ }\bibfield  {title} {\enquote {\bibinfo {title} {{Is timing
  noise important in the gravitational wave detection of neutron stars?}}}\
  }\href {\doibase 10.1103/PhysRevD.70.042002} {\bibfield  {journal} {\bibinfo
  {journal} {\prd}\ }\textbf {\bibinfo {volume} {70}},\ \bibinfo {eid} {042002}
  (\bibinfo {year} {2004})},\ \Eprint {http://arxiv.org/abs/0406045}
  {arXiv:0406045 [gr-qc]} \BibitemShut {NoStop}%
\bibitem [{\citenamefont {{Ashton}}\ \emph {et~al.}(2015)\citenamefont
  {{Ashton}}, \citenamefont {{Jones}},\ and\ \citenamefont
  {{Prix}}}]{ashton2015}%
  \BibitemOpen
  \bibfield  {author} {\bibinfo {author} {\bibfnamefont {G.}~\bibnamefont
  {{Ashton}}}, \bibinfo {author} {\bibfnamefont {D.~I.}\ \bibnamefont
  {{Jones}}}, \ and\ \bibinfo {author} {\bibfnamefont {R.}~\bibnamefont
  {{Prix}}},\ }\bibfield  {title} {\enquote {\bibinfo {title} {{Effect of
  timing noise on targeted and narrow-band coherent searches for continuous
  gravitational waves from pulsars}},}\ }\href {\doibase
  10.1103/PhysRevD.91.062009} {\bibfield  {journal} {\bibinfo  {journal}
  {\prd}\ }\textbf {\bibinfo {volume} {91}},\ \bibinfo {eid} {062009} (\bibinfo
  {year} {2015})},\ \Eprint {http://arxiv.org/abs/1410.8044} {arXiv:1410.8044
  [gr-qc]} \BibitemShut {NoStop}%
\bibitem [{\citenamefont {{Ashton}}\ \emph {et~al.}(2017)\citenamefont
  {{Ashton}}, \citenamefont {{Prix}},\ and\ \citenamefont
  {{Jones}}}]{ashton2017}%
  \BibitemOpen
  \bibfield  {author} {\bibinfo {author} {\bibfnamefont {G.}~\bibnamefont
  {{Ashton}}}, \bibinfo {author} {\bibfnamefont {R.}~\bibnamefont {{Prix}}}, \
  and\ \bibinfo {author} {\bibfnamefont {D.~I.}\ \bibnamefont {{Jones}}},\
  }\bibfield  {title} {\enquote {\bibinfo {title} {{Statistical
  characterization of pulsar glitches and their potential impact on searches
  for continuous gravitational waves}},}\ }\href {\doibase
  10.1103/PhysRevD.96.063004} {\bibfield  {journal} {\bibinfo  {journal}
  {\prd}\ }\textbf {\bibinfo {volume} {96}},\ \bibinfo {pages} {063004}
  (\bibinfo {year} {2017})},\ \Eprint {http://arxiv.org/abs/1704.00742}
  {arXiv:1704.00742 [gr-qc]} \BibitemShut {NoStop}%
\bibitem [{\citenamefont {{Prix}}(2009)}]{prix2009gravitational}%
  \BibitemOpen
  \bibfield  {author} {\bibinfo {author} {\bibfnamefont {R.}~\bibnamefont
  {{Prix}}},\ }\bibfield  {title} {\enquote {\bibinfo {title} {{Gravitational
  Waves from Spinning Neutron Stars}},}\ }in\ \href {\doibase
  10.1007/978-3-540-76965-1_24} {\emph {\bibinfo {booktitle} {Astrophysics and
  Space Science Library}}},\ \bibinfo {series} {Astrophysics and Space Science
  Library}, Vol.\ \bibinfo {volume} {357},\ \bibinfo {editor} {edited by\
  \bibinfo {editor} {\bibfnamefont {W.}~\bibnamefont {{Becker}}}}\ (\bibinfo
  {year} {2009})\ p.\ \bibinfo {pages} {651},\ \bibinfo {note}
  {\url{https://dcc.ligo.org/LIGO-P060039/public}}\BibitemShut {NoStop}%
\bibitem [{\citenamefont {{Jaranowski}}\ \emph {et~al.}(1998)\citenamefont
  {{Jaranowski}}, \citenamefont {{Kr{\'o}lak}},\ and\ \citenamefont
  {{Schutz}}}]{jks1998}%
  \BibitemOpen
  \bibfield  {author} {\bibinfo {author} {\bibfnamefont {P.}~\bibnamefont
  {{Jaranowski}}}, \bibinfo {author} {\bibfnamefont {A.}~\bibnamefont
  {{Kr{\'o}lak}}}, \ and\ \bibinfo {author} {\bibfnamefont {B.~F.}\
  \bibnamefont {{Schutz}}},\ }\bibfield  {title} {\enquote {\bibinfo {title}
  {{Data analysis of gravitational-wave signals from spinning neutron stars:
  The signal and its detection}},}\ }\href {\doibase
  10.1103/PhysRevD.58.063001} {\bibfield  {journal} {\bibinfo  {journal}
  {\prd}\ }\textbf {\bibinfo {volume} {58}},\ \bibinfo {eid} {063001} (\bibinfo
  {year} {1998})},\ \Eprint {http://arxiv.org/abs/gr-qc/9804014}
  {gr-qc/9804014} \BibitemShut {NoStop}%
\bibitem [{\citenamefont {{Edwards}}\ \emph {et~al.}(2006)\citenamefont
  {{Edwards}}, \citenamefont {{Hobbs}},\ and\ \citenamefont
  {{Manchester}}}]{edwards2006}%
  \BibitemOpen
  \bibfield  {author} {\bibinfo {author} {\bibfnamefont {R.~T.}\ \bibnamefont
  {{Edwards}}}, \bibinfo {author} {\bibfnamefont {G.~B.}\ \bibnamefont
  {{Hobbs}}}, \ and\ \bibinfo {author} {\bibfnamefont {R.~N.}\ \bibnamefont
  {{Manchester}}},\ }\bibfield  {title} {\enquote {\bibinfo {title} {{TEMPO2, a
  new pulsar timing package - II. The timing model and precision estimates}},}\
  }\href {\doibase 10.1111/j.1365-2966.2006.10870.x} {\bibfield  {journal}
  {\bibinfo  {journal} {\mnras}\ }\textbf {\bibinfo {volume} {372}},\ \bibinfo
  {pages} {1549--1574} (\bibinfo {year} {2006})},\ \Eprint
  {http://arxiv.org/abs/astro-ph/0607664} {astro-ph/0607664} \BibitemShut
  {NoStop}%
\bibitem [{\citenamefont {{Prix}}\ and\ \citenamefont
  {{Krishnan}}(2009)}]{prix2009}%
  \BibitemOpen
  \bibfield  {author} {\bibinfo {author} {\bibfnamefont {R.}~\bibnamefont
  {{Prix}}}\ and\ \bibinfo {author} {\bibfnamefont {B.}~\bibnamefont
  {{Krishnan}}},\ }\bibfield  {title} {\enquote {\bibinfo {title} {{Targeted
  search for continuous gravitational waves: Bayesian versus maximum-likelihood
  statistics}},}\ }\href {\doibase 10.1088/0264-9381/26/20/204013} {\bibfield
  {journal} {\bibinfo  {journal} {\cqg}\ }\textbf {\bibinfo {volume} {26}},\
  \bibinfo {eid} {204013} (\bibinfo {year} {2009})},\ \Eprint
  {http://arxiv.org/abs/0907.2569} {arXiv:0907.2569 [gr-qc]} \BibitemShut
  {NoStop}%
\bibitem [{\citenamefont {{Prix}}\ \emph {et~al.}(2011)\citenamefont {{Prix}},
  \citenamefont {{Giampanis}},\ and\ \citenamefont {{Messenger}}}]{prix2011}%
  \BibitemOpen
  \bibfield  {author} {\bibinfo {author} {\bibfnamefont {R.}~\bibnamefont
  {{Prix}}}, \bibinfo {author} {\bibfnamefont {S.}~\bibnamefont {{Giampanis}}},
  \ and\ \bibinfo {author} {\bibfnamefont {C.}~\bibnamefont {{Messenger}}},\
  }\bibfield  {title} {\enquote {\bibinfo {title} {{Search method for
  long-duration gravitational-wave transients from neutron stars}},}\ }\href
  {\doibase 10.1103/PhysRevD.84.023007} {\bibfield  {journal} {\bibinfo
  {journal} {\prd}\ }\textbf {\bibinfo {volume} {84}},\ \bibinfo {eid} {023007}
  (\bibinfo {year} {2011})},\ \Eprint {http://arxiv.org/abs/1104.1704}
  {arXiv:1104.1704 [gr-qc]} \BibitemShut {NoStop}%
\bibitem [{\citenamefont {Gelman}\ \emph {et~al.}(2013)\citenamefont {Gelman},
  \citenamefont {Carlin}, \citenamefont {Stern}, \citenamefont {Dunson},
  \citenamefont {Vehtari},\ and\ \citenamefont {Rubin}}]{gelman2013bayesian}%
  \BibitemOpen
  \bibfield  {author} {\bibinfo {author} {\bibfnamefont {A.}~\bibnamefont
  {Gelman}}, \bibinfo {author} {\bibfnamefont {J.~B.}\ \bibnamefont {Carlin}},
  \bibinfo {author} {\bibfnamefont {H.~S.}\ \bibnamefont {Stern}}, \bibinfo
  {author} {\bibfnamefont {D.~B.}\ \bibnamefont {Dunson}}, \bibinfo {author}
  {\bibfnamefont {A.}~\bibnamefont {Vehtari}}, \ and\ \bibinfo {author}
  {\bibfnamefont {D.~B.}\ \bibnamefont {Rubin}},\ }\href@noop {} {\emph
  {\bibinfo {title} {Bayesian Data Analysis}}},\ \bibinfo {edition} {3rd}\ ed.\
  (\bibinfo  {publisher} {CRC Press},\ \bibinfo {year} {2013})\BibitemShut
  {NoStop}%
\bibitem [{\citenamefont {{Archibald}}\ \emph {et~al.}(2013)\citenamefont
  {{Archibald}}, \citenamefont {{Kaspi}}, \citenamefont {{Ng}}, \citenamefont
  {{Gourgouliatos}}, \citenamefont {{Tsang}}, \citenamefont {{Scholz}},
  \citenamefont {{Beardmore}}, \citenamefont {{Gehrels}},\ and\ \citenamefont
  {{Kennea}}}]{archibald2013}%
  \BibitemOpen
  \bibfield  {author} {\bibinfo {author} {\bibfnamefont {R.~F.}\ \bibnamefont
  {{Archibald}}}, \bibinfo {author} {\bibfnamefont {V.~M.}\ \bibnamefont
  {{Kaspi}}}, \bibinfo {author} {\bibfnamefont {C.-Y.}\ \bibnamefont {{Ng}}},
  \bibinfo {author} {\bibfnamefont {K.~N.}\ \bibnamefont {{Gourgouliatos}}},
  \bibinfo {author} {\bibfnamefont {D.}~\bibnamefont {{Tsang}}}, \bibinfo
  {author} {\bibfnamefont {P.}~\bibnamefont {{Scholz}}}, \bibinfo {author}
  {\bibfnamefont {A.~P.}\ \bibnamefont {{Beardmore}}}, \bibinfo {author}
  {\bibfnamefont {N.}~\bibnamefont {{Gehrels}}}, \ and\ \bibinfo {author}
  {\bibfnamefont {J.~A.}\ \bibnamefont {{Kennea}}},\ }\bibfield  {title}
  {\enquote {\bibinfo {title} {{An anti-glitch in a magnetar}},}\ }\href
  {\doibase 10.1038/nature12159} {\bibfield  {journal} {\bibinfo  {journal}
  {\nat}\ }\textbf {\bibinfo {volume} {497}},\ \bibinfo {pages} {591--593}
  (\bibinfo {year} {2013})},\ \Eprint {http://arxiv.org/abs/1305.6894}
  {arXiv:1305.6894 [astro-ph.HE]} \BibitemShut {NoStop}%
\bibitem [{\citenamefont {{Livingstone}}\ \emph {et~al.}(2009)\citenamefont
  {{Livingstone}}, \citenamefont {{Ransom}}, \citenamefont {{Camilo}},
  \citenamefont {{Kaspi}}, \citenamefont {{Lyne}}, \citenamefont {{Kramer}},\
  and\ \citenamefont {{Stairs}}}]{livingstone2009}%
  \BibitemOpen
  \bibfield  {author} {\bibinfo {author} {\bibfnamefont {M.~A.}\ \bibnamefont
  {{Livingstone}}}, \bibinfo {author} {\bibfnamefont {S.~M.}\ \bibnamefont
  {{Ransom}}}, \bibinfo {author} {\bibfnamefont {F.}~\bibnamefont {{Camilo}}},
  \bibinfo {author} {\bibfnamefont {V.~M.}\ \bibnamefont {{Kaspi}}}, \bibinfo
  {author} {\bibfnamefont {A.~G.}\ \bibnamefont {{Lyne}}}, \bibinfo {author}
  {\bibfnamefont {M.}~\bibnamefont {{Kramer}}}, \ and\ \bibinfo {author}
  {\bibfnamefont {I.~H.}\ \bibnamefont {{Stairs}}},\ }\bibfield  {title}
  {\enquote {\bibinfo {title} {{X-ray and Radio Timing of the Pulsar in 3C
  58}},}\ }\href {\doibase 10.1088/0004-637X/706/2/1163} {\bibfield  {journal}
  {\bibinfo  {journal} {\apj}\ }\textbf {\bibinfo {volume} {706}},\ \bibinfo
  {pages} {1163--1173} (\bibinfo {year} {2009})},\ \Eprint
  {http://arxiv.org/abs/0901.2119} {arXiv:0901.2119 [astro-ph.SR]} \BibitemShut
  {NoStop}%
\bibitem [{\citenamefont {{Prix}}(2007)}]{prix2007}%
  \BibitemOpen
  \bibfield  {author} {\bibinfo {author} {\bibfnamefont {R.}~\bibnamefont
  {{Prix}}},\ }\bibfield  {title} {\enquote {\bibinfo {title} {{Template-based
  searches for gravitational waves: efficient lattice covering of flat
  parameter spaces}},}\ }\href {\doibase 10.1088/0264-9381/24/19/S11}
  {\bibfield  {journal} {\bibinfo  {journal} {\cqg}\ }\textbf {\bibinfo
  {volume} {24}},\ \bibinfo {pages} {S481--S490} (\bibinfo {year} {2007})},\
  \Eprint {http://arxiv.org/abs/0707.0428} {arXiv:0707.0428 [gr-qc]}
  \BibitemShut {NoStop}%
\bibitem [{\citenamefont {{Wette}}\ and\ \citenamefont
  {{Prix}}(2013)}]{wette2013}%
  \BibitemOpen
  \bibfield  {author} {\bibinfo {author} {\bibfnamefont {K.}~\bibnamefont
  {{Wette}}}\ and\ \bibinfo {author} {\bibfnamefont {R.}~\bibnamefont
  {{Prix}}},\ }\bibfield  {title} {\enquote {\bibinfo {title} {{Flat
  parameter-space metric for all-sky searches for gravitational-wave
  pulsars}},}\ }\href {\doibase 10.1103/PhysRevD.88.123005} {\bibfield
  {journal} {\bibinfo  {journal} {\prd}\ }\textbf {\bibinfo {volume} {88}},\
  \bibinfo {eid} {123005} (\bibinfo {year} {2013})},\ \Eprint
  {http://arxiv.org/abs/1310.5587} {arXiv:1310.5587 [gr-qc]} \BibitemShut
  {NoStop}%
\bibitem [{\citenamefont {{Wette}}(2015)}]{wette2015}%
  \BibitemOpen
  \bibfield  {author} {\bibinfo {author} {\bibfnamefont {K.}~\bibnamefont
  {{Wette}}},\ }\bibfield  {title} {\enquote {\bibinfo {title}
  {{Parameter-space metric for all-sky semicoherent searches for
  gravitational-wave pulsars}},}\ }\href {\doibase 10.1103/PhysRevD.92.082003}
  {\bibfield  {journal} {\bibinfo  {journal} {\prd}\ }\textbf {\bibinfo
  {volume} {92}},\ \bibinfo {eid} {082003} (\bibinfo {year} {2015})},\ \Eprint
  {http://arxiv.org/abs/1508.02372} {arXiv:1508.02372 [gr-qc]} \BibitemShut
  {NoStop}%
\bibitem [{\citenamefont {{Ashton}}\ and\ \citenamefont
  {{Prix}}(2018)}]{ashton2018}%
  \BibitemOpen
  \bibfield  {author} {\bibinfo {author} {\bibfnamefont {G.}~\bibnamefont
  {{Ashton}}}\ and\ \bibinfo {author} {\bibfnamefont {R.}~\bibnamefont
  {{Prix}}},\ }\bibfield  {title} {\enquote {\bibinfo {title} {{Hierarchical
  multi-stage MCMC follow-up of continuous gravitational wave candidates}},}\
  }\href@noop {} {\bibfield  {journal} {\bibinfo  {journal} {ArXiv e-prints}\ }
  (\bibinfo {year} {2018})},\ \Eprint {http://arxiv.org/abs/1802.05450}
  {arXiv:1802.05450 [astro-ph.IM]} \BibitemShut {NoStop}%
\bibitem [{\citenamefont {Foreman-Mackey}(2016)}]{corner}%
  \BibitemOpen
  \bibfield  {author} {\bibinfo {author} {\bibfnamefont {Daniel}\ \bibnamefont
  {Foreman-Mackey}},\ }\bibfield  {title} {\enquote {\bibinfo {title}
  {corner.py: Scatterplot matrices in python},}\ }\href {\doibase
  10.21105/joss.00024} {\bibfield  {journal} {\bibinfo  {journal} {The Journal
  of Open Source Software}\ }\textbf {\bibinfo {volume} {24}} (\bibinfo {year}
  {2016}),\ 10.21105/joss.00024}\BibitemShut {NoStop}%
\bibitem [{\citenamefont {{Christensen}}\ \emph {et~al.}(2004)\citenamefont
  {{Christensen}}, \citenamefont {{Dupuis}}, \citenamefont {{Woan}},\ and\
  \citenamefont {{Meyer}}}]{christensendupuis2004}%
  \BibitemOpen
  \bibfield  {author} {\bibinfo {author} {\bibfnamefont {N.}~\bibnamefont
  {{Christensen}}}, \bibinfo {author} {\bibfnamefont {R.~J.}\ \bibnamefont
  {{Dupuis}}}, \bibinfo {author} {\bibfnamefont {G.}~\bibnamefont {{Woan}}}, \
  and\ \bibinfo {author} {\bibfnamefont {R.}~\bibnamefont {{Meyer}}},\
  }\bibfield  {title} {\enquote {\bibinfo {title} {{Metropolis-Hastings
  algorithm for extracting periodic gravitational wave signals from laser
  interferometric detector data}},}\ }\href {\doibase
  10.1103/PhysRevD.70.022001} {\bibfield  {journal} {\bibinfo  {journal}
  {\prd}\ }\textbf {\bibinfo {volume} {70}},\ \bibinfo {eid} {022001} (\bibinfo
  {year} {2004})},\ \Eprint {http://arxiv.org/abs/gr-qc/0402038}
  {gr-qc/0402038} \BibitemShut {NoStop}%
\bibitem [{\citenamefont {Veitch}(2007)}]{veitch2007}%
  \BibitemOpen
  \bibfield  {author} {\bibinfo {author} {\bibfnamefont {J.~D.}\ \bibnamefont
  {Veitch}},\ }\emph {\bibinfo {title} {Applications of Markov Chain Monte
  Carlo methods to continuous gravitational wave data analysis}},\ \href
  {http://theses.gla.ac.uk/35/1/2007VeitchPhD.pdf} {Ph.D. thesis},\ \bibinfo
  {school} {University of Glasgow} (\bibinfo {year} {2007})\BibitemShut
  {NoStop}%
\bibitem [{\citenamefont {{Umst{\"a}tter}}\ \emph {et~al.}(2004)\citenamefont
  {{Umst{\"a}tter}}, \citenamefont {{Meyer}}, \citenamefont {{Dupuis}},
  \citenamefont {{Veitch}}, \citenamefont {{Woan}},\ and\ \citenamefont
  {{Christensen}}}]{umstatter2004}%
  \BibitemOpen
  \bibfield  {author} {\bibinfo {author} {\bibfnamefont {R.}~\bibnamefont
  {{Umst{\"a}tter}}}, \bibinfo {author} {\bibfnamefont {R.}~\bibnamefont
  {{Meyer}}}, \bibinfo {author} {\bibfnamefont {R.~J.}\ \bibnamefont
  {{Dupuis}}}, \bibinfo {author} {\bibfnamefont {J.}~\bibnamefont {{Veitch}}},
  \bibinfo {author} {\bibfnamefont {G.}~\bibnamefont {{Woan}}}, \ and\ \bibinfo
  {author} {\bibfnamefont {N.}~\bibnamefont {{Christensen}}},\ }\bibfield
  {title} {\enquote {\bibinfo {title} {{Estimating the parameters of
  gravitational waves from neutron stars using an adaptive MCMC method}},}\
  }\href {\doibase 10.1088/0264-9381/21/20/008} {\bibfield  {journal} {\bibinfo
   {journal} {\cqg}\ }\textbf {\bibinfo {volume} {21}},\ \bibinfo {pages}
  {S1655--S1665} (\bibinfo {year} {2004})}\BibitemShut {NoStop}%
\bibitem [{\citenamefont {{Abbott}}\ \emph {et~al.}(2010)\citenamefont
  {{Abbott}}, \citenamefont {{Abbott}}, \citenamefont {{Acernese}},
  \citenamefont {{Adhikari}}, \citenamefont {{Ajith}}, \citenamefont {{Allen}},
  \citenamefont {{Allen}}, \citenamefont {{Alshourbagy}}, \citenamefont
  {{Amin}}, \citenamefont {{Anderson}},\ and\ \citenamefont
  {et~al.}}]{abbott2010}%
  \BibitemOpen
  \bibfield  {author} {\bibinfo {author} {\bibfnamefont {B.~P.}\ \bibnamefont
  {{Abbott}}}, \bibinfo {author} {\bibfnamefont {R.}~\bibnamefont {{Abbott}}},
  \bibinfo {author} {\bibfnamefont {F.}~\bibnamefont {{Acernese}}}, \bibinfo
  {author} {\bibfnamefont {R.}~\bibnamefont {{Adhikari}}}, \bibinfo {author}
  {\bibfnamefont {P.}~\bibnamefont {{Ajith}}}, \bibinfo {author} {\bibfnamefont
  {B.}~\bibnamefont {{Allen}}}, \bibinfo {author} {\bibfnamefont
  {G.}~\bibnamefont {{Allen}}}, \bibinfo {author} {\bibfnamefont
  {M.}~\bibnamefont {{Alshourbagy}}}, \bibinfo {author} {\bibfnamefont {R.~S.}\
  \bibnamefont {{Amin}}}, \bibinfo {author} {\bibfnamefont {S.~B.}\
  \bibnamefont {{Anderson}}}, \ and\ \bibinfo {author} {\bibnamefont
  {et~al.}},\ }\bibfield  {title} {\enquote {\bibinfo {title} {{Searches for
  Gravitational Waves from Known Pulsars with Science Run 5 LIGO Data}},}\
  }\href {\doibase 10.1088/0004-637X/713/1/671} {\bibfield  {journal} {\bibinfo
   {journal} {\apj}\ }\textbf {\bibinfo {volume} {713}},\ \bibinfo {pages}
  {671--685} (\bibinfo {year} {2010})},\ \Eprint
  {http://arxiv.org/abs/0909.3583} {arXiv:0909.3583 [astro-ph.HE]} \BibitemShut
  {NoStop}%
\bibitem [{\citenamefont {Goggans}\ and\ \citenamefont
  {Chi}(2004)}]{goggans2004}%
  \BibitemOpen
  \bibfield  {author} {\bibinfo {author} {\bibfnamefont {P.~M.}\ \bibnamefont
  {Goggans}}\ and\ \bibinfo {author} {\bibfnamefont {Y.}~\bibnamefont {Chi}},\
  }\bibfield  {title} {\enquote {\bibinfo {title} {Using thermodynamic
  integration to calculate the posterior probability in bayesian model
  selection problems},}\ }\href {\doibase 10.1063/1.1751356} {\bibfield
  {journal} {\bibinfo  {journal} {AIP Conference Proceedings}\ }\textbf
  {\bibinfo {volume} {707}},\ \bibinfo {pages} {59--66} (\bibinfo {year}
  {2004})}\BibitemShut {NoStop}%
\bibitem [{\citenamefont {Skilling}(2006)}]{skilling2006nested}%
  \BibitemOpen
  \bibfield  {author} {\bibinfo {author} {\bibfnamefont {John}\ \bibnamefont
  {Skilling}},\ }\bibfield  {title} {\enquote {\bibinfo {title} {Nested
  sampling for general bayesian computation},}\ }\href {\doibase
  10.1214%2F06-BA127} {\bibfield  {journal} {\bibinfo  {journal} {Bayesian
  analysis}\ }\textbf {\bibinfo {volume} {1}},\ \bibinfo {pages} {833--859}
  (\bibinfo {year} {2006})}\BibitemShut {NoStop}%
\bibitem [{\citenamefont {{Allison}}\ and\ \citenamefont
  {{Dunkley}}(2014)}]{allison2014}%
  \BibitemOpen
  \bibfield  {author} {\bibinfo {author} {\bibfnamefont {R.}~\bibnamefont
  {{Allison}}}\ and\ \bibinfo {author} {\bibfnamefont {J.}~\bibnamefont
  {{Dunkley}}},\ }\bibfield  {title} {\enquote {\bibinfo {title} {{Comparison
  of sampling techniques for Bayesian parameter estimation}},}\ }\href
  {\doibase 10.1093/mnras/stt2190} {\bibfield  {journal} {\bibinfo  {journal}
  {\mnras}\ }\textbf {\bibinfo {volume} {437}},\ \bibinfo {pages} {3918--3928}
  (\bibinfo {year} {2014})},\ \Eprint {http://arxiv.org/abs/1308.2675}
  {arXiv:1308.2675 [astro-ph.IM]} \BibitemShut {NoStop}%
\bibitem [{\citenamefont {{Ashton}}\ and\ \citenamefont
  {{Keitel}}(2018)}]{pyfstat}%
  \BibitemOpen
  \bibfield  {author} {\bibinfo {author} {\bibfnamefont {G.}~\bibnamefont
  {{Ashton}}}\ and\ \bibinfo {author} {\bibfnamefont {D.}~\bibnamefont
  {{Keitel}}},\ }\href {\doibase 10.5281/zenodo.1243931} {\enquote {\bibinfo
  {title} {{PyFstat-v1.2}},}\ } (\bibinfo {year} {2018}),\ \bibinfo {note}
  {\url{https://doi.org/10.5281/zenodo.1243931}}\BibitemShut {NoStop}%
\end{thebibliography}%

\end{document}